# The gravitational waves are fictitious entities - II

A. LOINGER

Dipartimento di Fisica, Università di Milano

Via Celoria, 16,  20133 Milano, Italy

ABSTRACT. − Further arguments on the physical unreality of the gravitational waves of any kind.

**1.** − In a previous paper with the same title [1] I have shown that the gravitational waves of general relativity are non-physical sinuosities generated, in the last analysis, by undulating reference frames. I wish now to expound two additional, very straightforward arguments in favour of this thesis.

**2.** − Let us recall some things that the Monge-Hamilton theory of differential equations tells us about the pseudo-Riemannian geometry of general relativity [2], [3]. For any world point $x^s$ ($s$= 0, 1, 2, 3) we have a hypercone $C$ of vertex $x^s$, whose equation is

(2.1) $$g_{jk} (X^j - x^j) (X^k - x^k) = 0 \ , \ (j, k = 0, 1, 2, 3) \ ,$$

where the $X$'s are running co-ordinates; $C$ is a null hypercone. The set of all the null hypercones is a four-dimensional cone field described by the Mongean differential equation

(2.2) $$g_{jk} \ \frac{\mathrm{d}x^j}{\mathrm{d}p} \ \frac{\mathrm{d}x^k}{\mathrm{d}p} \ = \ 0 \ ,$$

− if $p$ is a suitable parameter −, and by the Hamiltonian differential equation

(2.3)
$$H := \tfrac{1}{2} \, g^{ik} \, \frac{\partial z}{\partial x^j} \, \frac{\partial z}{\partial x^k} = 0 \ ,$$

where $z = z \, ( \, x^0, x^1, x^2, x^3 )$. The (common) characteristic lines of (2.2) and (2.3) are the geodesic null lines. The set of all the null geodesics coming out from a given world point $a^s$ ($s = 0, 1, 2, 3$) generates a three-dimensional point manifold, which was called by Hilbert [2] the *Zeitscheide* belonging to $a^s$: it has in $a^s$ a nodal point, its tangent cone is the null cone belonging to $a^s$. If we write the equation of the *Zeitscheide* in the form

(2.4)
$$x^0 = \varphi \, ( \, x^1 , x^2 , x^3 ) \ ,$$

then $z = x^0 - \varphi \, ( \, x^1, x^2, x^3 )$ is a solution of (2.3).

**3.** – The functions $z \, (x^0, x^1, x^2, x^3)$ of the characteristics hypersurfaces $z = 0$ of d'Alembert equation are solutions of eq. (2.3), which is also the differential equation of the characteristics of Einstein field equation [3]. In general, $z \, (x^0, x^1, x^2, x^3) = 0$ gives the law of motion of a wave front, typically of an *electromagnetic* wave front.

Let us consider a hypothetical source $S$ of gravitational waves, which is practically concentrated in a space point $a^\sigma$ ($\sigma = 1, 2, 3$). (An assumption of this kind is at the basis of the classical theory of electrodynamical multipoles [4]). Assume now that $S$ begins to give out gravitational waves at some instant $a^0$. **In the wave zone** the space-time figure of the wave front (i.e. of the *Zeitscheide* belonging to $a^s$) will coincide with the corresponding portion of the surface of the null hypercone of vertex $a^s$. Now, a constant metric can always be impressed on the surface of a hypercone, and thus the imaginary nature of the gravitational waves becomes quite evident.

**4.** – An extreme, but conceptually expressive case is given by the $ds^2$ of Schwarzschild solution of a fixed point mass *M*. In the customary Droste-Weyl co-ordinates, we have

$$(4.1) \quad ds^2 = (1 - \frac{2\alpha}{r})\, c^2 dt^2 - (1 - \frac{2\alpha}{r})^{-1}\, dr^2 - r^2(d\vartheta^2 + \sin^2\vartheta\, d\varphi^2) \quad,$$

where $\alpha := GM/c^2$. Of course, (4.1) is physically significant *only for r* >2α, because, in particular: *i*) for *r* <2α the $ds^2$ is non-static (and non-reversible); *ii*) for *r* <2α the spatial co-ordinate *r*, which has been *defined* so that $4\pi K^2$ is the area of the surface *r = K*, would become a temporal co-ordinate. Other arguments are given in [5]. (Remark that the *original* $ds^2$ of the paper by Schwarzschild [6] is perfectly well-behaved in the entire space-time, with the obvious exception of the singular spatio-temporal line *r* = 0. An English translation of this fundamental memoir would powerfully contribute to ridicule the huge mass of senseless works about the imaginary notion of black hole).

Assume that $d\theta = d\varphi = 0$; then the condition $ds^2 = 0$ (null geodesics) implies that

$$(4.2) \qquad\qquad \frac{dr}{c\,dt} = \pm\,\frac{r - 2a}{r} \quad.$$

On the other hand, if $z = ct - \varphi(r)$ is the function of the characteristic hypersurface $z = 0$, it is

$$(4.3) \qquad 2H = \frac{r}{r - 2\alpha}\left(\frac{\partial z}{c\,\partial t}\right)^2 - \frac{r - 2\alpha}{r}\left(\frac{\partial z}{\partial r}\right)^2 =$$

$$= \frac{r}{r - 2a} - \frac{r - 2a}{r}\left(\frac{d\varphi}{dr}\right)^2 = 0 \qquad;$$

and therefore

(4.4) $$\varphi(r) = \pm \ (r - r_* + 2\alpha \ \ln \frac{r - 2\alpha}{r_* - 2\alpha}), \qquad ( \ r_* > 2\alpha).$$

Thus the characteristic hypersurface has the equation

(4.5) $$ct = \pm \ (r - r_* + 2\alpha \ \ln \frac{r - 2\alpha}{r_* - 2\alpha}) \ \ ,$$

from which it follows

(4.6) $$\frac{\mathrm{d}r}{c\mathrm{d}t} = - \frac{\partial z}{c\partial t} \Big/ \frac{\partial z}{\partial r} = \pm \ \frac{r - 2\alpha}{r} \ \ ,$$

i.e. again the result (4.2).

**5.** – Consider now a spinning top $E$, e.g. the Earth with respect to the system $F$ of the "fixed" stars. The believers in the existence of the gravitational radiation think that $E$ does emit gravitational waves. But if this were true, we should be obliged to conclude that *spatium est absolutum*, since $E$ would rotate in an absolute sense and $F$ would be motionless. Similarly, the source $S$ of sect. **3** and all the accelerated frames would be physically privileged systems. A conclusion which is evidently absurd from the standpoint of the exact (nonlinearized) formulation of the general theory of relativity.

\*————————————\*

∗─────────────────────────────∗